# Results from the ANTARES Neutrino Telescope

**Giorgio Giacomelli, for the Antares Collaboration**
University of Bologna and INFN Sezione di Bologna
e-mail: giacomelli@bo.infn.it

**Abstract.** The ANTARES underwater neutrino telescope is located in the Mediterranean Sea about 40 km from Toulon at a depth of 2475 m. In its 12 line configuration it has almost 900 photomultipliers in 275 "floors". The performance of the detector is discussed and several results are presented, including the measurements of downgoing muons, search for a diffuse flux of high energy muon neutrinos, search for cosmic point sources of neutrinos, search for fast magnetic monopoles, etc. A short discussion is also made on Earth and Sea Science studies with a neutrino telescope.

## I. INTRODUCTION

The effort to build large sea water Cherenkov detectors was pioneered by the Dumand Collaborattion with a prototype at great depths close to the Hawaii islands [1]; the project was eventually cancelled. Then followed the fresh water lake Baikal detector at relatively shallow depths [2]. Considerable progress was made by the AMANDA and IceCube telescopes in Antarctica [3,4]. In the Mediterranean Sea the NESTOR collaboration tested a deep line close to the Greek coast [5] and the NEMO Collaboration tested a number of prototypes close to Sicily [6]. The European groups are involved in a major project, Km3Net, aimed to the construction of a km$^3$ detector in the Mediterranean Sea [7].

ANTARES is a deep sea neutrino telescope, designed for the detection of high-energy neutrinos emitted by astrophysical sources, galactic and extragalactic [8]. The telescope is also sensitive to neutrinos produced via dark matter annihilation inside massive bodies like the Sun and the Earth. Other physics topics include the searches for fast magnetic monopoles, slow nuclearites, etc. ANTARES studies neutrinos from the southern hemisphere, which includes neutrinos from the center of our Galaxy. It is thus complementary to the studies by the ice detectors at the Southern Pole [4].

ANTARES is also a unique deep-sea marine observatory, providing continuous monitoring with a variety of sensors dedicated to oceanographic and Earth Science studies.

## II. THE ANTARES DETECTOR

The ANTARES neutrino telescope is located in the Mediterranean Sea, about 40 km from the coast of Toulon, France, at 42°48'N, 6°10'E, and at a depth of 2475 m [9] [10]. A schematic view of the detector is shown in Fig. 1. The detector is an array of photomultiplier tubes (PMTs) arranged on 12 flexible lines. Each line has up to 25 detection storeys of triplets of optical modules (OMs), each equipped with three downward looking 10 inch photomultipliers oriented at 45° to the line axis [11] [12] [13]. The lines are kept vertical by a buoy at the top of the 450 m long line. The spacing between storeys is 14.5 m and the lines are spaced by 60-70 m and are on an approximate octagonal structure [14]. An acoustic positioning system provides real time location of the detector elements with a precision of few centimeters. A system of optical beacons allows in situ calibrations. Since the end of 2008 the detector is running in its final configuration of 12 lines. An additional line (IL07) contains oceanographic sensors dedicated to the measurement of the environmental parameters. Line 12 and IL07 have hydrophone storeys to make calibrations and to measure the ambient acoustic backgrounds.

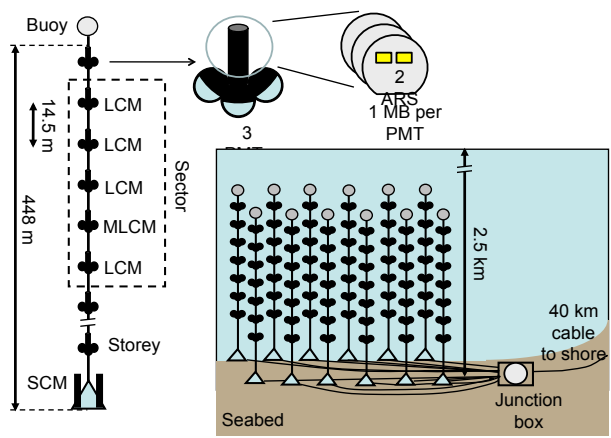

**Fig. 1**. The layout of the 12 line ANTARES detector. The sketches at the left and at the top show some details.

The lines are connected to the Junction Box (JB) that distributes power and data from/to shore. The instrumented part of each line starts at 100 m above the sea floor, so that Cherenkov light can be seen also from upgoing muons coming from neutrino interactions in the rock below the sea or in the sea water beneath the PMTs. The three dimensional structure of PMTs allows to measure the arrival time and position of Cherenkov photons produced by relativistic charged particles in the sea water. A reconstruction algorithm determines the direction of a muon, infer that of the incident neutrino and allows to distinguish upgoing muons, produced by neutrinos, from the more abundant downgoing muons, produced by cosmic ray interactions in the atmosphere.

The data acquisition is based on the "all data to shore" concept: all hits above a threshold of 0.3 single photoelectrons are digitized and sent to shore, where a computer farm applies a trigger requiring the presence of few coincidences between pairs of PMTs within a storey [15] [16]. This typical trigger rate is 5-10 Hz, dominated by downgoing muons.

For neutrino energies above few TeV the angular resolution for the search of astrophysical point sources is determined by the timing resolution and accuracy of the location of the photomultipliers. The relative time calibration is performed by a number of systems: the measurement of the transit time of the clock signals to the electronics in each storey and the determination of the residual time offsets within each storey. The positions of the PMts are measured by an acoustic positioning system to a precision of few cm [17] and checked by an optical beacon system [18]. The energy measurement relies on an accurate calibration of the charge detected by each PMT.

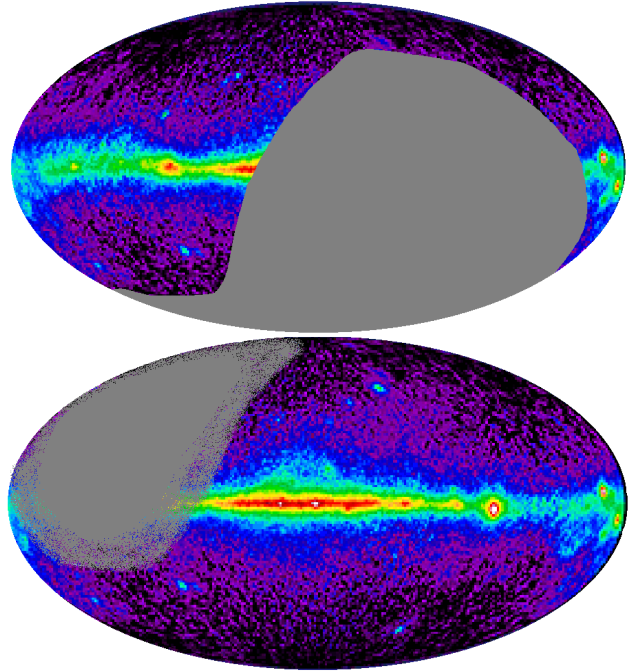

Fig. 2. Top: Regions of sky observable by neutrino telescopes at the South Pole (Amanda, IceCube), Bottom: in the Mediterranean Sea (ANTARES at 42° North). (Galactic coordinates).

### III. ATMOSPHERIC MUONS

The main signal observed by ANTARES is due to downgoing atmospheric muons, whose flux exceeds that of neutrino induced muons by many orders of magnitude. They are produced by high energy cosmic rays interacting with atomic nuclei of the upper atmosphere producing charged pions and kaons, which then decay into muons and neutrinos.

Atmospheric muons are used to check the detector response and to test different Monte Carlo simulations. They are also important to measure the attenuation of the muon flux as a function of depth and obtain the muon vertical depth intensity [19, 20, 21], and to verify the pointing accuracy of the detector (via the shadow of the cosmic ray flux by the moon [8]).

Fig. 3 bottom shows one muon bundle reconstructed by the ANTARES reconstruction algorithm. Fig. 4 shows the ANTARES muon vertical intensity data [21] together with a compilation of previous data [4,5] [22]. Fig. 8 shows a preliminary measurement of the Moon shadow. More extensive measurements by many past experiments were useful for establishing their pointing accuracy [23].

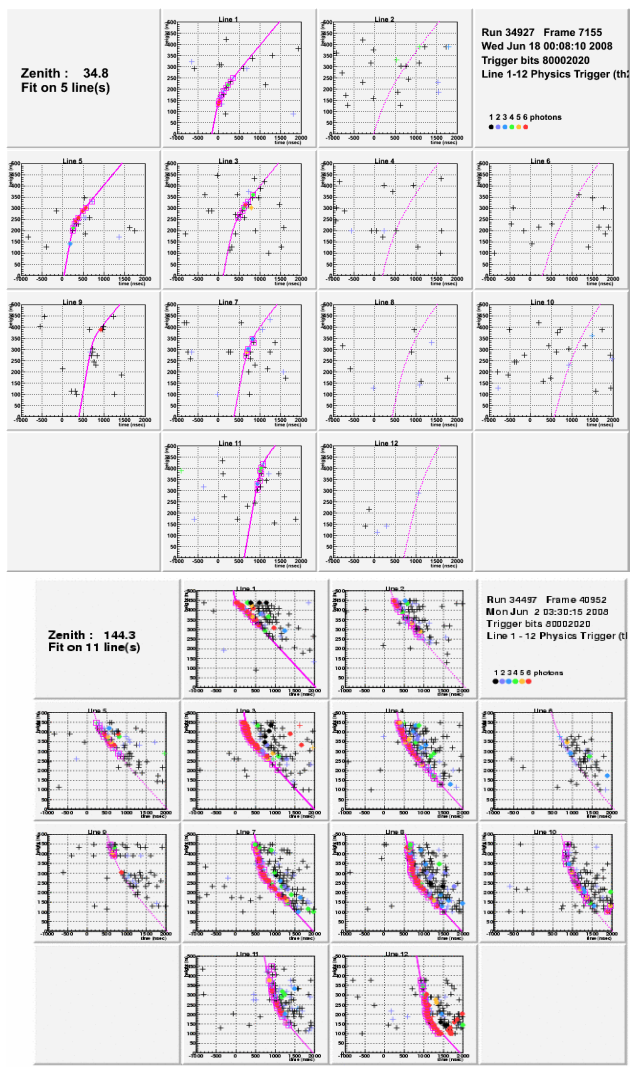

Fig. 3. Example of a neutrino candidate event seen in 5 ANTARES lines (top) and of a muon bundle seen in 11 lines (bottom).

## IV. NEUTRINO ASTRONOMY. COSMIC POINT SOURCES.

The main reason to build neutrino telescopes is to study high energy muon neutrino astronomy. Neutrinos may be produced in far away sources, where charged and neutral pions are produced and decay. The neutrinos from charged pion decay reach the Earth, while the *photons* from $\pi^0$ may interact with the Cosmic Microwave Background (CMB) radiation and with matter, *protons* are deflected by magnetic fields and *neutrons* are unstable. The main drawback of *neutrinos* is that one needs very large detectors.

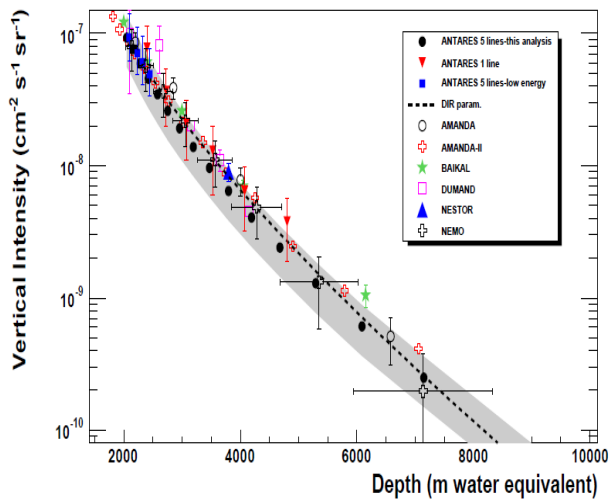

Fig. 4. Compilation of vertical muon intensity data of atmospheric muons vs equivalent water depth.

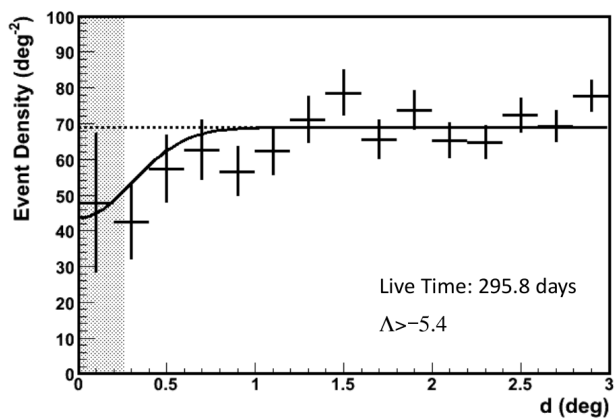

Fig. 5. Preliminary ANTARES muon data on the moon shadow.

The muons produced in neutrino interactions can be distinguished from atmospheric muons requiring that they are upgoing. Fig. 3 top shows one neutrino event which can be distinguished from a downgoing muon (bottom). Subsequent probability cuts allow to improve the separation. On the basis of the 2007+2008 data using a probability cut at Λ>−5.5 one has a neutrino sample with which to make a sky map in galactic coordinates, as shown in Fig. 5 [8]. Most of these events are from atmospheric muon neutrinos, which are an unavoidable background. A study is being made to measure this component and to obtain more information on atmospheric neutrinos [24]. The angular resolution (presently 0.5°) can be still improved by a precise timing [25] and the use of all the 12 lines.

A search amongst a predefined list of most promising cosmic sources found in γ ray astronomy can be made. Unfortunately with present statistics we do not find any confirmation with neutrinos above our present background. We can therefore obtain only 90% CL limits as shown in Fig. 6.

## V. HIGH ENERGY DIFFUSE NEUTRINO FLUX

The data were collected from December 2007 till December 2009 for a total livetime of 334 days. The runs had periods with 9, 10 or 12 lines in operation. Selected runs had a baseline <120 kHz and a burst fraction <40% [29]. The reconstruction algorithm yielded approximately 5% downgoing muons reconstructed as upgoing muons. They were discarded by applying geometry considerations and with a cut in the reconstruction quality parameter Λ>-5.5.

A neutrino energy estimator was used; it is based on the hit repetitions in the OMs due to the different arrival time of *direct* and *delayed* photons. The mean number of repetitions in the event is defined as the number of hits in the same OM within 500 ns from the earliest hit selected by the reconstruction algorithm : $R=R_i/N_{OM}$, where $R_i$ is the number of repetitions in OM i (in most cases $R_i$ is equal to 1, 2), $N_{OM}$ is the number of OMs in which hits used by the reconstruction algorithm are present. Fig. 7 shows the mean number of repetitions R versus the true MC neutrino-induced muon energy. Note that R is approximately linear with log $E_\mu$ in the logaritmic range 3.8-6.

The number of observed events after the proper

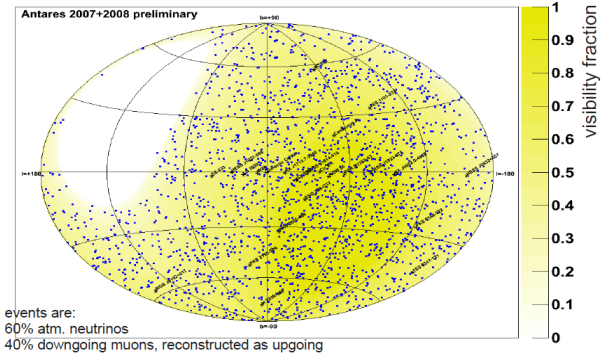

Fig. 5. Sky map in galactic coordinates of the upgoing neutrino candidates for the 2007+2008 data.

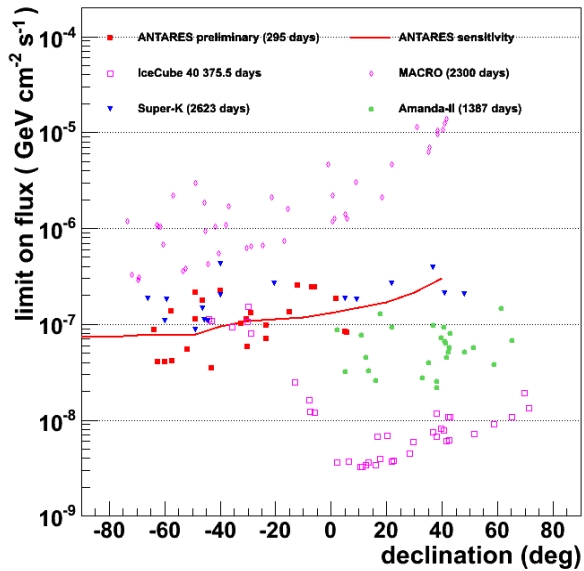

Fig. 6. 90% CL neutrino flux upper limits obtained by ANTARES (red dots) and other experiments (AMANDA [26], SuperK [27] and MACRO [28]).

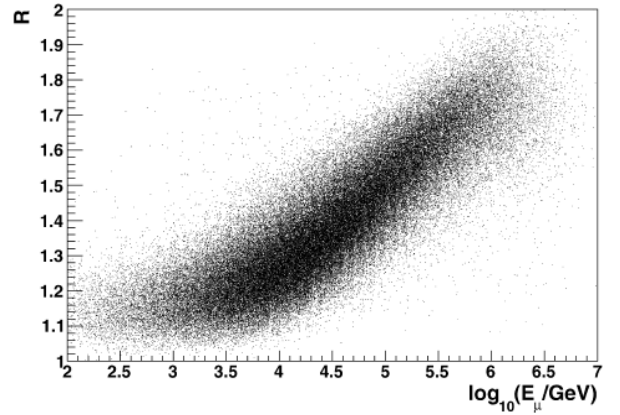

Fig. 7. Mean number of repetitions R as a function of the true neutrino-induced muon energy.

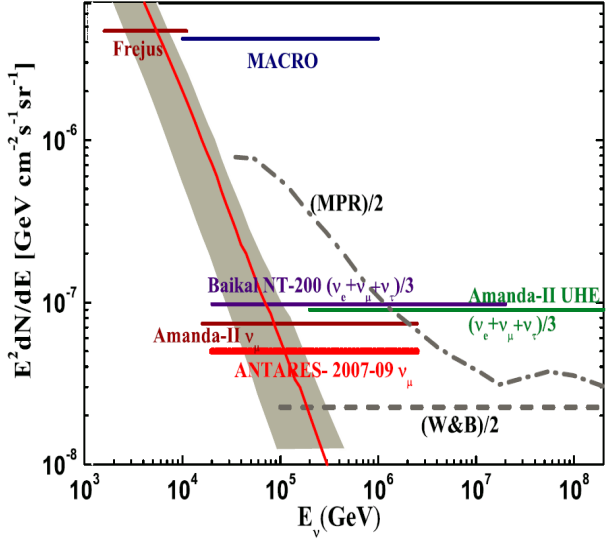

Fig. 8. The ANTARES 90% CL upper limit for a $E^{-2}$ diffuse high energy $\nu_\mu$+antiv$_\mu$ flux. The result is compared with the limits from other experiments. The grey band is the expected variation of the atmospheric $\nu_\mu$ flux and the central red line is averaged over all directions.

cuts is compatible with the expected background. One can thus only compute the 90% CL upper limit which at 90% CL for the quantity $E^2 \Phi$ is

$E^2 \Phi_{90\%}$=5.3 $10^{-8}$ cm$^{-2}$ s$^{-1}$ sr$^{-1}$

This limit is shown in Fig 8; the limit is valid in the energy range between 20 TeV to 2.5 PeV ; the limit is compared in Fig. 8 with other measured flux upper limits [30] and with two model predictions.

## VI. EXOTICA

Neutrino telescopes may also allow dark matter searches [31], searches for exotic particles (fast intermediate mass magnetic monopoles, slow nuclearites [32][33], etc.), searches for possible violations of general conservation laws [34], atmospheric neutrino studies [3][35], etc.

Magnetic Monopoles are required in many models of spontaneous symmetry breaking. The fast MMs which can be searched for in Neutrino Telescopes like ANTARES cannot be the GUT monopoles which are too massive, but could be the Intermediate Mass Magnetic monopoles with masses $10^{10}$-$10^{14}$ GeV [36]. Fig. 9 shows the preliminary limits obtained by ANTARES, which are compared with previous limits obtained by several experiments. In the limited beta range our limits compare well with those previously established [8]. And we can do considerably better with more data taking.

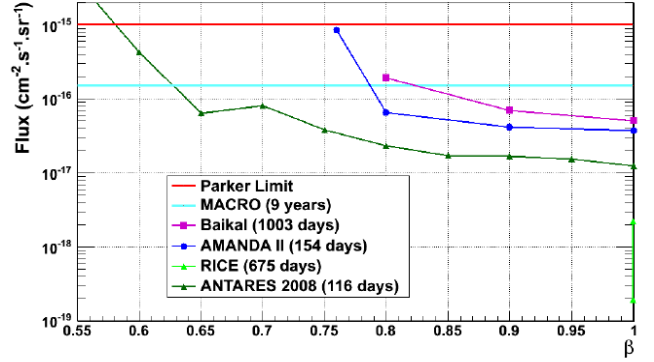

Fig. 9. Preliminary 90% CL limits on fast intermediate mass magnetic monopoles obtained by ANTARES. They are compared with previous limits obtained by AMANDA, Baikal and MACRO. The red horizontal line is the Parker Theoretical limit.

## VII. EARTH AND SEA SCIENCE

Earth and Sea Science is an interdisciplinary field with many important opportunities. ANTARES is well equipped for research in this field and the first preliminary results are being obtained now [8].

Fig. 10 shows the readings obtained by the underwater seismometer during the last great earthquake off the east coast of Japan (plus the terrible tsunami). The earthquake is well visible. Clearly the main purposes of an underwater seismometer are somewhat different, but one could say that one obtained a calibration of the instrument.

ANTARES is now equipped with an underwater camera which can show the luminous marine life, luminous bacteria and larger size luminous marine life, around the telescope.

In some of the ANTARES lines there is a very good instrumentation to measure with high precision the salinity and temperature of the bottom layer of the sea. And there are good instruments to measure the speed of the deep water currents, both in the horizontal plane and also in the vertical direction.

Preliminary analyses have indicated correlations of the light noise from luminous marine bacteria and from larger marine life with the sea currents, in particular also with vertical displacements of marine layers.

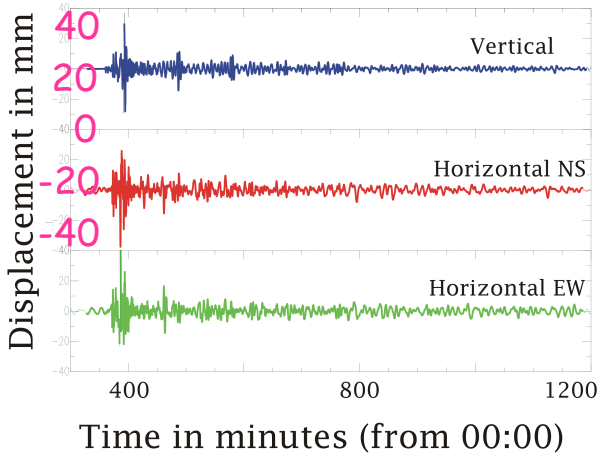

Fig. 10. The recent earthquake of extraordinary magnitude in Japan as observed by the seismometer at the ANTARES site.

VII. Conclusions. Outlook

The ANTARES neutrino telescope is running in its final 12 line configuration and is taking also data with a variety of instruments for Earth and Sea sciences.

The data on downgoing muons were used to calibrate the detector and to measure the vertical muon intensity as a function of the water depth. There is a reasonable agreement with previous data and with MC expectations. The larger part of the 30% systematic uncertainty is connected with a systematic uncertainty in the primary cosmic ray model [37].

First results of the search for cosmic point sources of neutrinos were obtained with an angular resolution of about 0.5°. We are now analyzing more data and will try to search for intermittent sources, in collaboration with optical telescopes [38].

Of particular interest is the study of neutrinos from the center of our galaxy.

A search was made for a diffuse flux of high energy neutrinos. Also this search will be repeated with more statistics.

The successful operation of ANTARES and the analyses results are important steps forward to Km3Net, a future deep sea km$^3$ neutrino observatory and marine sciences infrastructure planned for construction in the Mediterranean Sea.

**Acknowledgements.** I would like to acknowledge the cooperation, advice and technical support from many colleagues. In particular I thank S. Biagi, P. Coyle, S. Cecchini, A. Margiotta , M. Spurio and Y. Zornoza.